# Analyze Factors Influencing Drivers' Cell Phone Online Ride-hailing Software Using While driving: A Case Study in China

Xiangnan Song (1), Xianghong Li* (1), Kai Yin (2), Huimin Qi (1), Xufei Fang (1)

(1) Department of Transportation, Henan Polytechnic University, Jiaozuo 454003, China
(2) School of traffic and transportation, Beijing Jiaotong University, Beijing 100044, China
*Corresponding author.

**Abstract:** The road safety of traffic is greatly affected by the driving performance of online ride-hailing, which has become an increasingly popular travel option for many people. Little attention has been paid to the fact that the use of cell phone online ride-hailing software by drivers to accept orders while driving is one of the causes of traffic accidents involving online ride-hailing. This paper, adopting the extended theory of planned behavior, investigates the factors that factors influencing the behavior of Chinese online ride-hailing drivers cell phone ride-hailing software usage to accept orders while driving. Results showed that attitudes, subjective norms, and perceived behavioral control have a significant and positive effect on behavioral intentions. Behavioral intention is most strongly influenced by attitude. There is no direct and significant impact of group norms on behavioral intention. Nonetheless, group norms exert a substantial and beneficial influence on attitude, subjective norms, and perceived behavioral control. This study has discovered, through a mediating effect test, that attitude, subjective norm, and perceived behavioral control play a mediating and moderating role in the impact of group norm on behavioral intention. These findings can offer theoretical guidance to relevant departments in developing effective measures for promoting safe driving among online ride-hailing drivers.

## 1 Introduction

For a long time, distracted driving has been a worldwide issue and has been identified as one of the key factors contributing to road traffic accidents (Charlton, 2009; Lee et al., 2013; Nee et al., 2019; Razi-Ardakani et al., 2019). According to the US Automobile Federation Traffic Safety Foundation, distracted driving is when the driver's attention is diverted from the driving task by objects, people, or actions inside or outside the car, resulting in the driver's inability to respond effectively to potential hazards (Dewar and Olson, 2007). The definition attributes driver distraction to a divided attention. The driver's attention diversion is mainly concentrated on electronic devices, especially the use of cell phones. For instance, Redelmeier et al. (1997) conducted a study and found that the likelihood of a driver using a mobile phone and causing a traffic accident within 10 minutes prior to the accident was 4.3 times greater than that of a driver not using a mobile phone. Caney et al. (2016) discovered that distracted driving behaviors were present in more than 75% of rear-end collisions, with cell phone use, looking outside the vehicle, and talking to passengers being the most prevalent. Due to the demands of their jobs, online taxi drivers are at an elevated risk of distraction compared to other types of drivers, as they must be constantly on their cell phones. This unique work requirement makes them particularly vulnerable to distractions while driving. Barrios et al. (2019) found that the introduction of internet dating was associated with a 3% increase in accidents involving passengers and pedestrians in cars. And Nguyen-phuoc et al. (2020a) point out that the use of mobile phones is one of the key causes of traffic accidents in online ride-hailing. The above statement indicates the need to conduct research on the distracted behavior of drivers using mobile phones, especially among online ride-hailing drivers. However, this group has been rarely investigated in distracted driving-related research in China.

The drivers of online ride-hailing have a dual identity of ordinary drivers and operating drivers. Because of their specific identity, individuals in this group may experience different psychological changes than the average person, which can increase the likelihood of unsafe driving behavior. Among all the unsafe driving behaviors, the online car-hailing software is one of the main sources of distraction (Dang, 2017; Peng, 2023). The popularity of online ride-hailing software (He and Shen, 2015) and the rising incidence of accidents involving these services (Truong and Nguyen, 2019) have led to a growing research interest in studying the behavior of drivers who use these apps on their mobile phones.

Prior studies on the use of cell phone ride-hailing software by drivers have primarily aimed to establish models or techniques that can effectively reduce the driving risks associated with these apps (Ding et al., 2023; Tang et al., 2017; Tang et al., 2018). For example, Ding et al. (2023) investigated a general optimization method with the goal of improving the safety of ride-hailing services and mitigating the risk of drivers using mobile phones while driving. Tang et al. (2018) constructed a distraction detection model based on driving performance by conducting simulated driving experiments, which was used to identify visual and cognitive distractions as well as normal driving behaviors of drivers while using online ride-hailing software. Furthermore, studies have demonstrated that ride-hailing drivers face the highest risk when using mobile phones while driving (Nguyen-Phuoc et al., 2020a). Regarding the process of online ride-hailing drivers using cell phones to accept ride requests, Yun (2020) analyzed the unsafe behaviors during the order-taking process and subsequently explored the motivations behind these unsafe behaviors. The aforementioned studies indicate that there is a scarcity of research on the utilization of online software by ride-hailing drivers while operating a vehicle.

Many studies have focused on the increased risk of crashes caused by cell phone use while driving (Ishigami et al., 2009; Violanti et al., 1996; Dingus et al., 2016). For instance, Redelmeier and Redelmeier (1997) surveyed 699 drivers and analyzed their cell phone calls on the day of the collision and the previous week. They found that using cell phones in motor vehicles links to quadrupling the crash risk during the brief time interval involving a call. Violanti et al. (1996) also found that talking on a mobile phone for more than 50 minutes per month in the car increased the risk of a traffic accident by 5.59 times. In order to increase the safety level of road driving, some researchers have begun to study how to reduce the risk of collisions caused by using mobile phones while driving. (Mase et al., 2020; Molina et al., 2013; Wang et al., 2020). In addition, many researchers have used the theory of planned behavior to explore the factors influencing drivers' choice to use mobile phones while driving from a social psychological perspective (Hassani et al., 2017; Brandt et al., 2022; Montuori et al., 2021; Muladilijiang et al., 2022; Batoul et al., 2017). For example, Montuori et al. (2021) found that the knowledge held by the driver is not related to the behavior, and the attitude is closely related to the knowledge and behavior. Using a four-part questionnaire grounded in the theory of planned behavior, Batuol et al. (2017) surveyed 400 drivers and observed notable variations in the normative, control, and behavioral beliefs of drivers who engage in mobile phone use while driving. Drawing parallels with this concept, this paper builds upon the theory of planned behavior (TPB) (Ajzen, 1991) to investigate the underlying factors that drive online ride-hailing drivers to accept orders through cell phone online ride-hailing software while they are driving.

This study puts forward three key contributions. First of all, despite the confirmation that ride-hailing increases the risk of road safety (Barrios et al., 2019; Nguyen-Phuoc et al., 2020a), previous research has not yet incorporated group norms (GN) as an extended latent variable into the

TPB model of related research. Secondly, while there has been extensive research on the factors that influence distracted driving among drivers, there are limited studies on the same topic among online ride-hailing drivers, particularly in relation to the use of online software for accepting ride orders while driving. This study reveals the distracting behavior of online car-hailing drivers when using online car-hailing software to pick up orders. Thirdly, while scholars have extensively researched various factors that contribute to distracted driving and traffic accidents among online ride-hailing drivers, such as working hours (Rayle et al., 2014) and economic factors (Mao et al., 2021), there is a dearth of studies that specifically investigate the use of cell phone ride-hailing software for accepting orders while driving. This study adds to the existing literature by examining the behavior of Chinese online ride-hailing drivers in typical developing countries who use cell phone online ride-hailing software to accept orders while driving.

This study utilized an extended Theory of Planned Behavior (TPB) model to conduct a questionnaire survey and establish a structural equation model (SEM) for understanding the factors that influence online ride-hailing drivers' use of cell phone online ride-hailing software to accept orders while driving.  Gaining a comprehensive understanding of these factors can offer theoretical guidance for designing more focused interventions and enhancing road traffic safety in China.

## 2 Method

### 2.1 Theoretical foundation

2.1.1 Theory of Planned Behavior

In 1991, Ajzen proposed the Theory of Planned Behavior (Ajzen, 1991). The theory is employed to elucidate the process of decision-making in general behavior. Studies have confirmed its ability to predict certain behavioral intentions and actions of travelers in the transportation domain. The Theory of Planned Behavior has been implemented to investigate dangerous driving behaviors, such as speeding (Christie and Ward, 2019; Cristea et al., 2013), drunk driving (Potard et al., 2018), and driving while fatigued (Jiang et al., 2017). Moreover, it has been used with success to predict drivers' cell phone use while operating a vehicle. For example, Nemme et al. (2010) utilized the extended TPB to predict young drivers' intention to read and send text messages while driving, and the results indicate that TPB can be used to explain drivers' behavior of texting while driving. Qu et al. (2020) examined the influence of various WeChat functions on driving behavior by analyzing the self-reports of 286 drivers from China through the lens of the theory of planned behavior. The findings indicate that drivers' attitude is a significant predictor of their engagement in sending text messages, listening to voice messages, and sending and browsing pictures on WeChat while driving. Zhou et al. (2009) investigated the intention of young drivers to use handheld or hands-free mobile phones while driving from the perspective of the theory of planned behavior. The analysis revealed that TPB can account for 43% and 48% of the variance in the intention to use hands-free and handheld mobile phones while driving, respectively.

TPB suggests that personal intention has a direct impact on actual behavior, and the three cognitive factors of attitude (ATT), subjective norm (SN), and perceived behavioral control (PBC) collectively influence the behavioral intention (BI). Attitude pertains to an individual's positive or negative emotions towards a behavior, subjective norms relate to the social influence that prompts an individual to perform a behavior, and perceived behavioral control concerns the perceived difficulty of executing a behavior. Behavioral intention reflects an individual's assessment of the subjective likelihood of performing a behavior, which indicates their readiness to adopt the behavior (Ajzen, 1991; Ajzen and Driver, 1992; Fishbein and Ajzen, 1975). In the basic model of the theory of planned behavior, a more positive attitude, stronger subjective norms, and greater perceived behavioral control are associated with a stronger influence on behavior intention (Ajzen, 2002). Therefore, building on the same logic, this paper presents the following hypothesis:

H1: The online ride-hailing drivers' ATT toward using the online car-hailing software to receive orders while driving has a significantly positive effect on their BI to do so.

H2: The positive impact of SN on the BI of online car-hailing drivers to use online car-hailing software to take orders while driving is significant.

H3: Online car-hailing drivers' PBC has a significant positive effect on their BI to use online car-hailing software to take orders while driving.

TPB is considered a flexible theory by some researchers (Ajzen, 1991; Armitage and Conner, 2001). While the basic TPB model can account for a significant portion of behavioral intention, its predictive power can be improved by including additional factors. The incorporation of group norms (GN) as an additional variable has been shown to enhance the predictive power of behavioral intention (Terry et al., 1999; White et al., 2009). Therefore, this study adds an additional variable to the traditional TPB model: group norms.

Typically, group norms establish a shared standard that defines the collective preferences of group members and directs their actions (Hogg and Reid, 2006). Evidence suggests that group norms exert a meaningful influence on both behavior and intention, and can even augment the predictive power of the TPB model (Terry and Hogg, 1996; Walsh et al., 2009). Given this background, the following hypothesis is proposed in this paper:

H4: GN exert a notable positive influence on ATT;

H5: GN exert a notable positive influence on SN;

H6: GN exert a notable positive influence on PBC;

H7: GN exert a notable positive influence on BI.

2.1.2 Extended TPB Model

The aim of this study is to investigate the psychological factors underlying the driving distraction caused by the use of online car-hailing software to take orders while driving in China. Drawing on the TPB theoretical framework, this paper presents an extended TPB model that incorporates a new variable (GN) into the existing constructs of ATT, SN, PBC, and BI, in order to investigate the factors that drive Chinese online car-hailing drivers to use online car-hailing software to accept orders while driving. As a further contribution, this study investigates the influence of the GN variable on Chinese online ride-hailing drivers' intention to use cell phone online ride-hailing software to accept orders while driving, and examines the underlying processes that drive this relationship. Figure 1 depicts the extended TPB model proposed in this study.

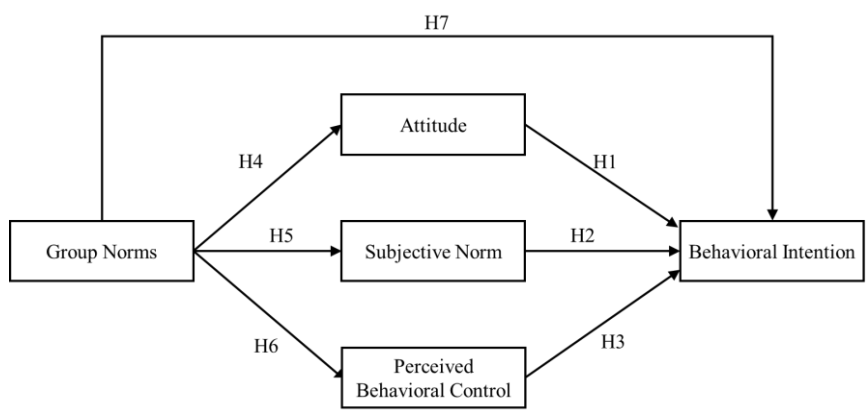

Fig 1. Extended TPB Model

## 2.2 Data preparation

2.2.1 Questionnaire survey

The questionnaire consists primarily of two sections. In the first part, the five constructs of the extended TPB model, namely ATT, SN, PBC, GN, and BI, are measured. A 5-point Likert scale was utilized for the measurement process. Participants who scored higher were more liable to have a positive response. Table 1 displays the questionnaire design, which encompassed a total of 15 items in its first part.

Table 1 The expanded TPB questionnaire design

| Constructs | Measure items | Response range |
|---|---|---|
| ATT | ATT1: For you, it is the right thing to accept ride-hailing orders on a mobile app while driving | (1) strongly disagree; (5) strongly agree |
| | ATT2: For you, using cell mobile ride-hailing software to accept orders while driving will not affect your normal driving | |
| | ATT3: For you, it is inevitable to operate the online ride-hailing software while driving | |
| SN | SN1: Your family or friends would hope you to use the online ride-hailing software to accept orders while driving | (1) strongly disagree; (5) strongly agree |
| | SN2: Your family member or friend agrees that you can use the online ride-hailing software to accept orders while driving | |
| | SN3: Your family or friends think you should use online ride-hailing software to accept orders while driving | |
| PBC | PBC1: Performing online ride-hailing software operations while driving, for you (degree of difficulty) | (1) very difficult; (5) very easy |
| | PBC2: Performing online ride-hailing software operations while driving, for you (level of danger) | (1) very dangerous; (5) very safe |
| | PBC3: Would you be able to multitask by driving and using a ride-hailing app competently? | (1) not at all;(5) complete line |
| GN | GN1: Considering your friends who are also online car-hailing drivers, how many people do you think would choose to accept orders on online ride-hailing software while driving? | (1) no one; (5) everyone |
| | GN2: Considering your friends who are also online car-hailing drivers, how many of them think it is safe to accept orders from online ride-hailing software while driving? | |
| | GN3: Does your platform or management support your " accept orders on online ride-hailing software while driving " behavior? | (1) not supported at all; (5) fully supported |
| BI | BI1: Do you wish to use online ride-hailing software to accept orders while driving? | (1) very undesired; (5) very much desired |
| | BI2: What is the likelihood that you will continuously use ride-hailing software to accept orders while driving? | (1)Definitely not; (5)Definitely |
| | BI3: In the future, do you still wish to accept ride-hailing orders on your mobile device while driving | (1) very undesired; (5) very much desired |

In the second part of the questionnaire, participants were asked to provide demographic data, including gender, age, highest level of education, monthly income, full-time or part-time driver status, and number of accidents in the past three years.

2.2.2 Participants

The final sample included 397 Chinese online ride-hailing drivers, including 319 males, accounting for 80.4%, and 78 females, accounting for 19.6%. Most participants were between 26 and 45 years old (63%). In a recent study, Jiang and Yang (2023) surveyed 375 online ride-hailing drivers, showing that most respondents were male (71.1%). There is not much difference between the full-time online ride-hailing drivers (44.6%) and part-time online ride-hailing drivers (55.4%) in this article, which is similar to the number of samples selected by Shi et al. (2023). In summary, the participants in this study can be considered representative of the ride-hailing driver population. In addition, it is worth noting that 168 drivers (42.3%) stated that the number of traffic accidents (including friction and minor collisions) in the past three years was 3-5.

### 2.3 Measures

SEM, which is a multivariate technique, has been extensively used in social and behavioral sciences for fitting and testing hypothetical models (Hai et al., 2023). To investigate the hypothesized relationship between latent variables in the extended TPB model, a SEM was constructed for online ride-hailing drivers who use cell phone online ride-hailing software while driving in this study. The first step in data analysis is to conduct Confirmatory Factor Analysis. The analysis of the measurement model in SEM includes both reliability and validity analysis. Once a satisfactory measurement model is obtained, the subsequent step is to estimate the path coefficients using the maximum likelihood estimation method. This process entails analyzing the structural model in SEM to confirm the accuracy of each hypothesis.

## 3 Result analysis

### 3.1 Descriptive statistical analysis

Table 2 displays the average values of statistical variables based on various social characteristics. In general, all of the latent variables received higher scores from men compared to women. This indicates that men generally have a more favorable assessment of online ride-hailing compared to women; however, it is worth noting that the sample size may have contributed to this result since there were more male drivers than female drivers in the sample.

This survey included participants who were between 18 and 65 years of age. The age groups in this paper are divided by a threshold of 35 years old, with individuals under 35 categorized as young and those over 35 categorized as middle-aged. Based on the statistics, it was discovered that there is minimal variation in the scores of the five latent variables between young and middle-aged individuals. This indicates that there is a relatively consistent perception across different age groups towards the behavior of accepting online ride-hailing orders while driving.

From an educational perspective, drivers with high school or technical secondary school education and above have higher scores in attitude, subjective norms, and behavioral intentions than those with high school or technical secondary school education and below. Compared to the latter group, the former group is more acknowledged for engaging in the behavior of accepting ride-hailing orders while driving, and their friends and loved ones also agree with them. Nevertheless, drivers who have completed high school or technical secondary school education hold the belief that utilizing cell phone online ride-hailing software while driving may lead to increased driving difficulty, which could be associated with their familiarity with the software.

As for income, since none of the collected samples had a monthly income below 3000 yuan, the monthly incomes were categorized based on the benchmark of 5000 yuan. The ratings of drivers earning more than 5000 yuan per month were higher than those earning less, which could be attributed to their monthly work hours. Since online ride-hailing drivers' income is calculated based on the number of orders, those with higher incomes usually have a larger total order volume, resulting in higher ratings for drivers earning over 5000 yuan.

It can be seen from the statistical data that part-time drivers score higher than full-time drivers on attitudes, subjective norms, and behavioral intentions, and full-time drivers score higher than part-time drivers on perceived behavioral control and group norms This result may be because full-time drivers generally have more extended driving experience and more skilled driving skills. Hence, they experience less difficulty driving a vehicle using online ride-hailing software. At the same time, part-time drivers' jobs involve various industries. At the same time, the work of part-time drivers involves various industries. Part-time online ride-hailing drivers have a lower sense of belonging to the industry than full-time online ride-hailing drivers. The former only use their spare time to work part-time as online ride-hailing drivers. It is worth noting that ride-hailing drivers who have had 3-5 accidents in the past three years have higher ratings in all aspects.

Table 2 Social demographic characteristics descriptive statistics

| Variables | | ATT | SN | PBC | GN | BI |
|---|---|---|---|---|---|---|
| Gender | Man | 4.03 | 3.92 | 3.82 | 3.91 | 4.12 |
| | Woman | 3.86 | 3.52 | 3.56 | 3.70 | 3.44 |
| Age | 35 years and under | 3.97 | 3.89 | 3.77 | 3.93 | 4.01 |
| | Over 35 years old | 4.03 | 3.79 | 3.78 | 3.82 | 3.96 |
| Education degree | High school or technical secondary school and below | 3.98 | 3.80 | 3.80 | 3.88 | 3.98 |
| | High school or technical secondary school and above | 4.04 | 3.92 | 3.72 | 3.86 | 4.00 |
| Monthly profit | 3000-5000 yuan | 3.99 | 3.72 | 3.75 | 3.68 | 3.85 |
| | More than 5000 yuan | 4.00 | 3.84 | 3.77 | 3.87 | 3.98 |
| Type of driver | Full-time drivers | 3.99 | 3.83 | 3.86 | 3.95 | 3.95 |
| | Part-time drivers | 4.01 | 3.85 | 3.70 | 3.81 | 4.02 |
| In 3 years, the number of accidents | 1-2 times | 3.95 | 3.78 | 3.77 | 3.86 | 3.91 |
| | 3-5 times | 4.05 | 3.93 | 3.79 | 3.91 | 4.01 |
| | More than 5 times | 4.00 | 3.84 | 3.77 | 3.87 | 3.98 |

**3.2 Measurement model**

Through the measurement model, the correlation between latent variables and items is analyzed. Confirmatory factor analysis is utilized in this paper to appraise the structure's reliability and validity, with validity testing consisting of both convergent and discriminant validity. The dependability of a survey is measured by its reliability, which is manifested in the consistency, reproducibility, and stability of its results. (Ling and Fang, 2003). Cronbach's α coefficient and comprehensive reliability (CR) were employed in this paper to examine the reliability of the questionnaire (Shah et al., 2023; Zhang et al., 2020). A Cronbach's alpha coefficient greater than 0.7 (Hair et al., 2011) and a CR greater than 0.6 (Hair et al., 2016) indicate good reliability. The α coefficient ranging from 0.769 to 0.862 and the CR value ranging from 0.771 to 0.875 are evidence of the scale's good reliability.

Convergent validity, also called aggregation validity, assesses the degree to which each item on a factor is highly correlated with the other items on that factor, indicating whether the items on the scale are tapping into the same construct. The present study employs average variance extraction (AVE) to assess the convergent validity of the project (Shah et al., 2023). The AVE value of the latent variable, which ranges from 0.529 to 0.667 as presented in Table 3, indicates that the convergence validity is satisfactory.

Discriminant validity, also referred to as differential validity, measures the extent to which a construct differs from other constructs (Hair et al., 2016). Attaining satisfactory discriminant validity requires that the correlation between each latent variable be below the square root of the AVE value of that variable. (Nguyen-Phuoc et al., 2020b). The results presented in Table 4 indicate that this questionnaire possesses good discriminant validity.

Table 3 Results of reliability and validity statistics

| Constructs | Items | Factor loadings | Cronbach's α | CR | AVE |
|---|---|---|---|---|---|
| ATT | ATT1 | 0.752 | 0.769 | 0.771 | 0.529 |
| | ATT2 | 0.726 | | | |
| | ATT3 | 0.704 | | | |
| SN | SN1 | 0.792 | 0.817 | 0.817 | 0.599 |
| | SN2 | 0.732 | | | |
| | SN3 | 0.796 | | | |
| PBC | PBC1 | 0.721 | 0.772 | 0.772 | 0.530 |
| | PBC2 | 0.719 | | | |
| | PBC3 | 0.744 | | | |
| GN | GN1 | 0.759 | 0.811 | 0.807 | 0.582 |
| | GN2 | 0.757 | | | |
| | GN3 | 0.772 | | | |
| BI | BI1 | 0.821 | 0.862 | 0.857 | 0.667 |
| | BI2 | 0.829 | | | |
| | BI3 | 0.799 | | | |

Table 4 Discriminant validity

| Constructs | GN | PBC | ATT | SN | BI |
|---|---|---|---|---|---|
| GN | 0.763 | | | | |
| PBC | 0.329 | 0.728 | | | |
| ATT | 0.407 | 0.134 | 0.723 | | |
| SN | 0.371 | 0.122 | 0.151 | 0.774 | |
| BI | 0.396 | 0.322 | 0.386 | 0.327 | 0.817 |

**3.3 Structural model**

To investigate the impact of behavioral attitudes, subjective norms, perceived behavioral control, and group norms on the intention of online car-hailing drivers to use online car-hailing software to accept orders while driving, this paper employs IBM SPSS AMOS 26 software to conduct a path analysis of the TPB-based structural model. The path coefficients are estimated using the maximum likelihood estimation method.

3.3.1 Path coefficient estimation

Table 5 displays the estimation of the path coefficients. The findings suggest that most of the proposed assumptions are validated. In this model, there is a significant positive correlation between the ATT, SN, and PBC of online car-hailing drivers and BI, which supports hypothesis 1, hypothesis 2, and hypothesis 3. Furthermore, Hypothesis 4, Hypothesis 5, and Hypothesis 6 are supported by the significant positive influence of GN on ATT, SN, and PBC. Hypothesis 7 cannot be supported based on the fact that GN do not have a direct positive impact on BI.

Table 5 Path coefficient estimation table

| Paths in Mode | Estimate | S.E. | C.R. | p-value | Hypotheses |
| --- | --- | --- | --- | --- | --- |
| BI ← ATT | 0.333 | 0.082 | 4.058 | *** | Supported |
| BI ← SN | 0.217 | 0.063 | 3.425 | *** | Supported |
| BI ← PBC | 0.237 | 0.068 | 3.494 | *** | Supported |
| BI ← GN | 0.149 | 0.072 | 2.067 | 0.039 | Not Supported |
| ATT ← GN | 0.353 | 0.057 | 6.206 | *** | Supported |
| SN ← GN | 0.384 | 0.066 | 5.799 | *** | Supported |
| PBC ← GN | 0.320 | 0.066 | 4.863 | *** | Supported |

Note: *** $p < 0.001$. The path coefficients shown in the table are standardized path coefficients.

### 3.3.2 Model fitting test

Various fitness indicators must be satisfied by the model to assess the fit of the data and model. Previous research has informed the selection of six fitness indicators for testing in this paper (Shen et al., 2020; Yun, 2020): chi-square with degrees of freedom ($\chi^2/df$), Root Mean Squared Error of Approximation (RMSEA), comparative fit index (CFI), incremental fit index (IFI), goodness-of-fit index (GFI) and Tucker-Lewis index (TLI). When the $\chi^2/df$ value is between 1 and 3, the RMSEA value is less than 0.08, and the CFI, IFI, GFI and TLI values are greater than 0.90, the fitting effect is considered to be good (Hu and Bentler, 1999). The values for each index in Table 6 meet their respective standards, indicating that the model fits the survey data well.

Table 6 The fitting index of the model

| Fit index | Measured value | Standard value |
| --- | --- | --- |
| $\chi^2/df$ | 1.546 | $1 < \chi^2/df < 3$ |
| RMSEA | 0.037 | $< 0.08$ |
| CFI | 0.980 | $> 0.9$ |
| IFI | 0.980 | $> 0.9$ |
| GFI | 0.956 | $> 0.9$ |
| TLI | 0.974 | $> 0.9$ |

### 3.3.3 Structural equation model

The SEM of online ride-hailing drivers using the cell phone online ride-hailing software to accept orders while driving, based on the extended TPB, is depicted in Figure 2. The direct impact on intention is greatest for ATT, followed by PBC and SN, while GN has no direct impact on BI. In addition, GN has the most significant positive effect on ATT, with SN and PBC following.

While GN does not have a direct effect on BI, it does have a direct and significant effect on ATT, which in turn has a direct effect on BI. Similarly, PBC and SN can be directly influenced by GN, and these two variables can in turn have a direct impact on BI. Thus, to investigate whether GN can have an indirect impact on BI, this paper employs the bootstrap analysis method. With a sample size of 5000, the main effect is statistically significant at the 95% confidence level. (Huang et al., 2021). This study examined the mediating effects of three paths: group norms → attitude → behavioral intention, group norms → subjective norms → behavioral intention, and group norms → perceived behavioral control → behavioral intention. Table 7 displays the results. The analysis revealed that the indirect effects of these three paths were statistically significant ($P < 0.05$) (Zhang and Li, 2020). According to the results, GN has no direct effect on BI, but it does have an indirect effect through ATT, SN, and PBC.

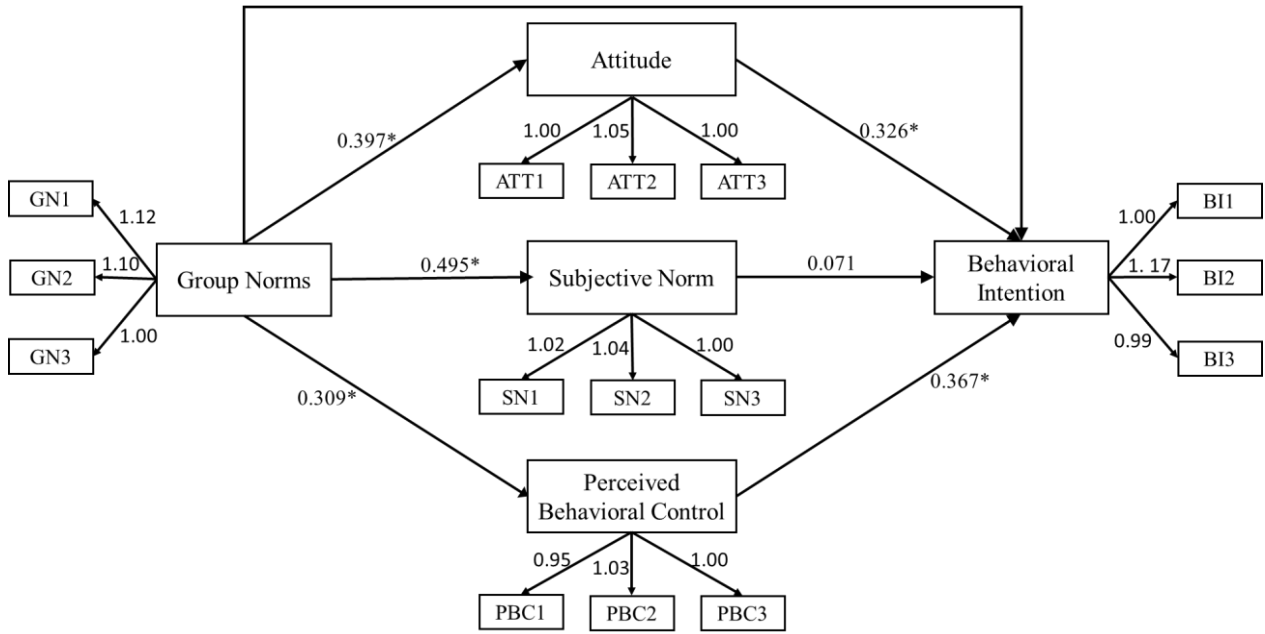

Fig 2. The SEM of the online ride-hailing drivers using the cell phone online ride-hailing software to accept orders while driving.(ATT1 represents correctness of behavior; ATT2 represents that normal driving will not be affected; ATT3 represents inevitable; SN1 represents that this is desired by family and friends; SN2 represents this was agreed to by family and friends; SN3 represents that your family and friends believe that you ought to do this; PBC1 represents level of difficulty; PBC2 represents level of peril; PBC3 represents level of ease; GN1 represents what percentage of people in your field do this as well; GN2 represents how many people in your profession think this is safe; GN3 represents management's view; BI1 represents willingness; BI2 represents possibility; BI3 represents approaching)

Table 7 Mediating effect test

| Path relationship | Effect type | Estimate | 95% confidence interval | | P |
|---|---|---|---|---|---|
| | | | Lower | Upper | |
| GN→ATT→BI | Indirect Effect | 0.109 | 0.045 | 0.210 | 0.001 |
| | Direct Effect | 0.139 | -0.009 | 0.294 | 0.063 |
| | Total Effect | 0.248 | 0.103 | 0.400 | 0.003 |
| GN→SN→BI | Indirect Effect | 0.078 | 0.028 | 0.148 | 0.003 |
| | Direct Effect | 0.139 | -0.009 | 0.294 | 0.063 |
| | Total Effect | 0.216 | 0.082 | 0.359 | 0.003 |
| GN→PBC→BI | Indirect Effect | 0.071 | 0.026 | 0.144 | 0.001 |
| | Direct Effect | 0.139 | -0.009 | 0.294 | 0.063 |
| | Total Effect | 0.209 | 0.064 | 0.369 | 0.006 |

### 3.4 Structural equation modeling analysis

To comprehend the factors that influence the use of cell phone online ride-hailing software by Chinese online ride-hailing drivers to accept orders while driving, we obtained relevant data through a questionnaire survey and developed an SEM based on the extended TPB model.

This paper finds that online ride-hailing drivers' intention to use cell phone online ride-hailing software to accept orders while driving is positively influenced by ATT, SN, and PBC, which is consistent with the research outcomes of some researchers (Huang et al., 2021; Waddell and Wiener, 2014). They are of the opinion that all the standard TPB factors hold substantial importance in determining BI. The findings of this study suggest that drivers' intention to use cell phone online ride-hailing software to accept orders while driving is likely to increase with positive attitudes, strong subjective norms, and higher perceived behavioral control.

Consistent with the findings of numerous scholars, attitude is identified as the most potent direct predictor of behavioral intention (Gauld et al., 2017; Walsh et al., 2008). They are of the opinion that attitude is the strongest determinant in anticipating the intention to use cell phones while driving. Online ride-hailing drivers' stance on accepting orders while driving reveals their perception that such a practice can lead to more time efficiency and profit gains. The trust that passengers have in online ride-hailing drivers is directly proportional to their trust in the ride-hailing platform, as per a study (Sun and Elefteriadou, 2011). As the driver's score on the platform increases, so does the passengers' likability towards them. Ride-hail drivers must accept more orders during their working hours to obtain higher scores on the platform. Despite being aware of the high risk of danger associated with using a mobile phone while driving, many drivers still choose to engage in this behavior, and this can be explained by the aforementioned reasons (Shi et al., 2016).

The impact of PBC on BI is second only to that of ATT. PBC is an indicator of the confidence one has in their ability to successfully execute and control a particular behavior (Ajzen, 2002). The study found that perceived behavioral control had a significant impact on the behavioral intention of online car drivers. This suggests that these drivers do not face difficulty in simultaneously driving and accepting online orders, and they exhibit a high level of confidence in their ability to perform both tasks.

The significant impact of SN on BI indicates that family and friends can influence the behavior of online ride-hailing drivers. If online ride-hailing drivers' behavior is acknowledged by significant family or friends, they are more inclined to engage in such behavior. As Yun (2020) proposed, traffic safety education should be targeted not only towards drivers, but also towards their families and friends, as a result. It's noteworthy that, in comparison to ATT and PBC, SN has a weaker direct influence on BI.

In addition, GN have no significant effect on BI. The findings indicate that the high GN observed in this study is not a strong predictor of drivers' intention to use cell phone online ride-hailing software while driving. This contradicts the results reported by Qu et al. (2020) and Narges et al. (2019). A reasonable explanation for this result is that many online ride-hailing drivers are part-time. They come from various industries and do not have a strong sense of professional belonging to the side job of online ride-hailing drivers. Moreover, the application threshold for online ride-hailing driver status is relatively low. After a driver joins an online ride-hailing company or online ride-hailing platform, the company or platform lacks centralized management and restraint on online ride-hailing drivers. For some behaviors that may cause driving distraction, for example, drivers cannot operate mobile phones while driving, there are no clear and mandatory punitive measures, and management is relatively evasive.

Table 7 reveals an interesting finding that, while GN does not have a direct significant impact on BI, it can still influence BI indirectly through three paths: group norms → attitude → behavioral intention, group norms → subjective norms → behavioral intention, and group norms → perceived behavioral control → behavioral intention. The structural equation model shows that GN have the greatest impact on ATT. When friends who work in the same industry as online car-hailing drivers approve of the practice of accepting driving orders, the attitude of online car-hailing drivers will become more positive, which in turn increases the likelihood of their engagement in such behavior. The effect of GN on BI is moderated by SN and PBC. Rewritten sentence: Supportive feedback from friends and family, coupled with drivers' confidence in multitasking driving and accepting orders, leading to a more robust intention to drive and utilize cell phone online ride-hailing software. Rewritten sentence: In general, while group norms do not have a significant direct impact on behavioral intention, they can still have an indirect effect on behavioral intention by influencing attitude, subjective norms, and perceived behavioral control in TPB. This highlights the need to enhance the group norms of online ride-hailing drivers.

## 4 Research prospects and limitations

By utilizing the extended theory of planned behavior to create a structural equation model for accepting online ride-hailing orders while driving, this study contributes to the literature and offers practical significance for enhancing the safety of online ride-hailing drivers. Nonetheless, the study has limitations:

Firstly, when introducing new latent variables, other latent variables that may also influence behavioral intentions, such as moral norms (Nemme and White, 2010), knowledge (Montuori et al., 2021), and anticipated regret (Gauld et al., 2014), were not taken into account. Additional variables can be introduced in future research to enhance the predictive capability of the model.

Secondly, the purpose of this experiment is to comprehensively enhance the driving safety of online ride-hailing drivers by analyzing the psychological factors that influence the use of cell phone online ride-hailing software while driving. Driving simulation experiments (Hibberd et al., 2013) or road measurement (Hancock et al., 2003) can be used in the future to provide online ride-hailing drivers with a technically feasible intervention scheme.

Thirdly, the relationship between demographic variables and behavioral intention is not studied in this paper. Studies conducted previously have established a significant link between demographics and the practice of using mobile phones while driving (Truong and Nguyen, 2019). Therefore, additional research is necessary to explore the impact of demographics on this conduct.

Finally, while BI is regarded as a reliable predictor of behavior (Ajzen, 1991), subsequent studies have revealed that the predictive power of intention on behavior may diminish depending on the research methodology employed (Jiang et al., 2016). Hence, the subsequent investigation ought to incorporate an examination of the association between BI and behavior.

## 5 Conclusions

The study offers a comprehensive understanding of the behavior displayed by drivers of online car-hailing services who use online car-hailing software to accept orders while driving. The structural equation model of online ride-hailing drivers utilizing cell phone online ride-hailing software to accept orders while driving is developed using the extended theory of planned behavior. The primary findings are summarized below:

(1) The strongest direct influence on BI is attributed to ATT, indicating that it is the most critical factor that affects drivers' use of cell phone online ride-hailing software to accept orders while driving. The correlation between attitude and intention demonstrates that as the ATT towards using cell phone online ride-hailing software to accept orders while driving becomes more positive, the BI of online ride-hailing drivers to engage in this behavior also increases.

(2) PBC exerts a significant positive influence on BI. The ease with which drivers can use cell phone ride-hailing software to accept orders directly affects their behavior.

(3) SN also has a significant and positive impact on BI. For online ride-hailing drivers, the opinions of their friends and family have a direct impact on their behavioral intention to use cell phone online ride-hailing software to accept orders while driving.
(4) While GN do not directly and significantly impact BI, they do have a positive and significant impact on ATT, SN, and PBC. As GN increase, ride-hailing drivers are more inclined to have a positive attitude towards using cell phone online ride-hailing software while driving to accept ride requests. Moreover, their friends and relatives are more likely to endorse this behavior, and drivers' confidence in their ability to handle both tasks simultaneously will be higher.
(5) The influence of GN on BI is mediated by ATT, SN, and PBC.

These findings can serve as a theoretical foundation for concerned organizations to develop efficient measures that enhance the safety of ride-hailing drivers and minimize their unsafe behavior.